Zr and Mo macrosegregation in Ti6246 titanium alloy industrial-scale ingot by vacuum arc remelting

S.X.Zhu[a,b], Q.J.Wang[a,b]*, J.R.Liu[b], Z.Y.Chen[b]

[a] School of Materials Science and Engineering, University of Science and Technology of China, Shenyang 11016, China

[b] Shi Changxu Innovation Center for Advanced Materials, Insititude of Metal Research, Chinese Academy of Sciences，Shenyang 110016, China

Abstract

Zr and Mo macrosegregations were investigated in Ti6246 titanium alloy industrial-scale ingot (Φ720 mm × 1160 mm) by vacuum arc remelting. The formation mechanism of Zr and Mo macrosegregations was studied during the solidification process. Zr macrosegregation was characterized by low content in the equiaxed grain zone and high content in the hot top zone. Mo exhibits an opposite trend with Zr. The macrosegregations of Zr and Mo were the most pronounced, with a statistic segregation degree higher than Al and Sn. It could be concluded that temperature gradient and solidification rate dominated the macrosegregation formations during the solidification process. The thermal buoyancy made the negative segregation Zr be continuously discharged to the front of the solid-liquid interface. Mo was enriched in the solid phase at the solid-liquid interface as the positive segregation.

## 1. Introduction

Near-α or α+β titanium alloys such as Ti-6-2-4-2S(Ti-6Al-2Sn-4Zr-2Mo), Ti6246(Ti-6Al-2Sn-4Zr-6Mo) and Ti-17(Ti-5Al-2Sn-2Zr-4Mo-4Cr) are excellent candidates for aerospace applications [1]. With the increase of Si, Sn or Zr and decrease of Mo, among all explored near-α titanium alloys, IMI834 (Ti-5.5Al-4.0Sn-3.5Zr-0.5Mo-0.7Nb-0.35Si-0.06C) [2], Ti-1100 (Ti-6Al-2.7Sn-4Zr-0.4Mo-0.45Si) [3], Ti60 (Ti-5.8Al-4.0Sn-3.5Zr-0.4Mo-0.4Nb-1.0Ta-0.4Si-0.06C)

[4] alloys have been used up to a maximum service temperature of about 600 ℃. Aluminum (Al), tin (Sn), zirconium (Zr), and molybdenum (Mo) are mainly alloying elements in all the above-mentioned alloys . Compared with Ti, Al and Sn are lower melting-point and density elements and will be volatilized during vacuum arc remelting(VAR) progress. Zr is a slightly higher melting-point and boiling point, grave higher density than Ti. Mo has a very high melting point up to 2623 ℃, about 1000 ℃ higher than Ti, and twice density than Ti. So, difficulties in controlling the chemical homogeneities on the macro- and micro-scales have been well known.

Titanium alloys exhibit many problems at the billet evaluation stage which are connected to the melting and casting processes used in their preparation [5-7]. Mitchell [5] concluded that β flecks are connected with the melting practice. Conventional melting by double or triple VAR gives a product whose reliability in these defects is uncomfortably close to the limits imposed by probabilistic design and fracture mechanics analysis. Dobatkin et al. [6] revealed that positive zonal segregation appears in titanium alloy ingots under the conditions of vacuum arc remelting. Positive segregation also occurs in aluminum alloy ingots during vacuum arc remelting or continuous casting accompanied by vigorous metal stirring in a molten pool. Various relationships of these parameters govern the development of a type of macrosegregation (positive or negative) or production of a rather homogeneous ingot. Liu et al. [7] reported a novel method for the analysis of composition distribution of titanium alloys over a large area (64 mm ×72 mm) was investigated by exploring the original position statistic distribution based on spark spectrum (OPA-SS) in Ti-6.5Al-1.0Cr-0.5Fe-6.0Mo-3.0Sn-4.0Zr titanium alloy. The macrosegregation of Sn was the most pronounced, with a statistic segregation degree higher than 18%; the macrosegregation of Mo followed with a statistic segregation degree of 10%; the macrosegregation of Al and Fe was relatively milder, lower than 8%. The main reason for the macrosegregation state of the as-cast Ti-6.5Al-1.0Cr-0.5Fe-6.0Mo-3.0Sn-4.0Zr alloy can be the solute redistribution during liquid solidification and the diffusion rate of each element in the solid phase. However, there are no introductions about the Zr segregation in that article. Although the details of the segregation in the solidification in remelting process have been discussed [8], many of the works were to analyze the smaller ingot less Φ380 mm [6,7], and focus on the significant quantities of β stabilizing elements such as Fe and Cr in particular and "hard-α" (LDI) and high density inclusions (HDI) in microsegregation. Some works just were to model the macrosegregation during the vacuum arc remelting [9]. The conclusions are still controversial and few experimental evidences have been published on the segregation behaviors of element macrosegregation in industry scale titanium alloy ingot. In Ti6246 alloy, the bulk composition contains 6%Al，2%Sn, 4%Zr and 6% Mo (wt.%, the same below). Lower melting-point and density elements (Al and Sn) maybe are not the origin of metallurgical quality problems, but higher melting-point and density elements (Zr and Mo) have not been investigated in detail as a source of segregation.

This article aims to illustrate the segregation behavior of alloying elements and explain the formation mechanisms of Zr and Mo macrosegregation in Ti6246 alloy industrial-scale ingot.

## 2. Material and experimental

The Ti6246 alloy ingot used in the experiment was melted by triple VAR with the actual production composition Ti-6.15Al-2.05Sn-4Zr-6Mo. Traditional technologies of VAR, i.e., the operations of electromagnetic stirring on melt pool and hot top set points at the end of the VAR process were used. While other parameters are listed in Table 1, to represent the actual value for industry scale ingot. The ingot of size Φ720 mm × 1160 mm was machined with a smooth surface by milling and the components of five locations in the surface were analyzed from the top to the bottom. The ingot was cut along the longitudinal direction to analyze the longitudinal section. The schematics of the ingot cutting and analytical sampling positions are shown in Fig.1.

**Table 1**: Main parameters of VAR for Ti6246 alloy

| Parameters | First remelting | Second remelting | Third remelting |
|---|---|---|---|
| Diameter of crucible(mm) | Φ550 | Φ620 | Φ720 |
| Pre-vacuum(⩽Pa) | 3 | 2 | 2 |
| Leakage rate(⩽Pa/min) | 0.93 | 0.7 | 0.7 |

| Vacuum in stable melting stage(≤Pa) | 7 | 5 | 5 |
|---|---|---|---|
| electromagnetic stirring current(A) | 18(direct) | 15(alternating)/20S | 10(alternating)/2S |
| Voltage in stable melting stage(V) | 22–37 | 23–39 | 23–40 |
| Current in stable melting stage(kA) | 14–16 | 22–25 | 23–28 |

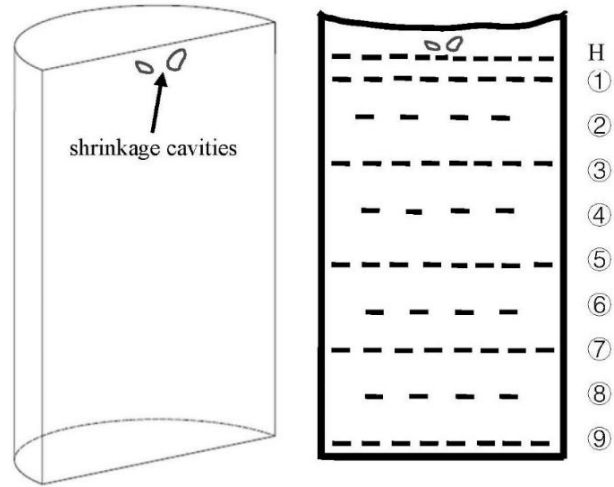

**Fig. 1** The schematics of the ingot cutting and analytical sampling positions

After observation of the solidification macrostructure along the longitudinal section, seven billets, which were cut along the longitudinal with the thickness 30 mm were heated at a special scheme of 940 ℃/1 h/Water Quenching(WQ). The β transformation temperature ($T_β$) of Ti6246 is about 960 ℃ determined by the metallographic method. The ($T_β$-20 ℃)/1 h/WQ is for the convenience of detecting β-flecks using an optical microscope (OM), which is a normal operation for observation of β-flecks in Ti-17 alloy [10].

For structure observations, all samples were prepared by the standard mechanical polishing operation. They were subsequently etched in a mixture solution of HF:$HNO_3$:$H_2O$ with a ratio of 1:2:50 for macro/microstructure observation. The macrostructure was examined on a digital camera. The microstructure was examined on a scanning electron microscope (SEM, STEMI2000). The components in the different areas of the ingot were analyzed using inductive coupled plasma emission spectrometer (ICP) method.

## 3. Results

3.1 Macrostructure of the ingot

Fig. 2 shows the longitudinal macrostructure of the 2.1 ton Ti6246 ingot. No clear beta flecks or high density inclusions were found in different positions of the ingot. However, macrostructure heterogeneity was found, which is characterized by three zones. Three zones refer to columnar grain zone, equiaxed grain zone and hot top zone. The columnar grain zone appears at the round of the ingot including the bottom, widths are about between 110 mm in the top and 300 mm in the bottom. There is only a tiny equiaxed grain zone left in the surface of the ingot formed by the rapid solidification rate because the ingot of size Φ720 mm was machined into a smooth surface with a size Φ680 mm by milling. So the tiny equiaxed grain zone in the surface of the ingot can be considered to be the origin of the columnar grain zone. The columnar grain zone is about 40% of the ingot, and the columnar grain grows along the direction to two obvious shrinkage cavities in the hot top zone. There is a boundary between the equiaxed grain zone and the columnar grain zone. The equiaxed grain zone was in the shape of "circular truncated cone", the diameters of the circular truncated cone were about 460 mm on the top and 340 mm at the bottom. The equiaxed grain zone in the middle of the ingot is about 60% of the ingot including most of the hot top, which is similar as 60%–70% of the total ingot volume, which is still liquid at the time the process sequence enters into the hot-top cycle with the low solidification rate for equiaxed grain [8]. The hot top zone is on the top of the ingot, with two obvious shrinkage cavities in the center. There is almost equiaxed grain near the shrinkage cavities in the hot top center except a few columnar grains formed on the top of the ingot.

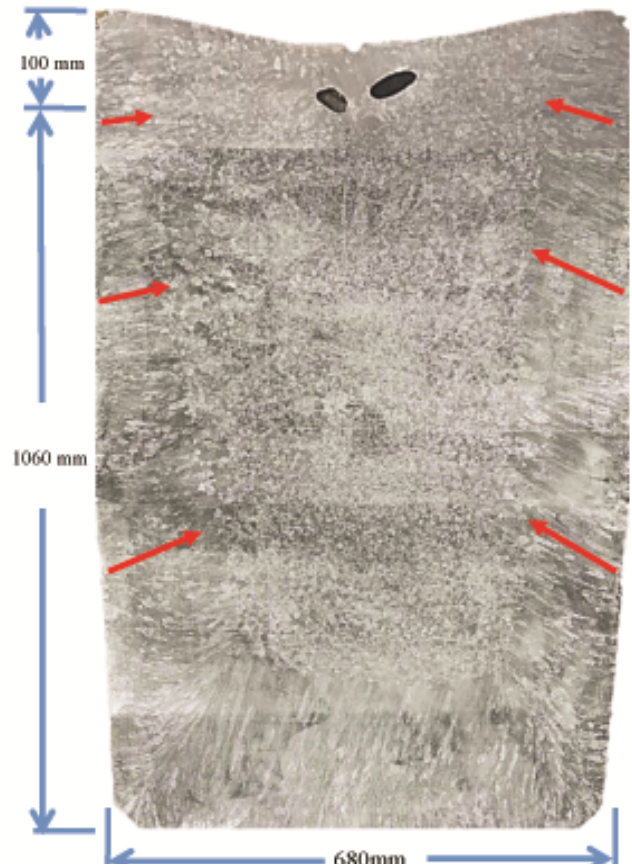
Fig. 2 macrostructure of Ti6246 ingot

3.2 Effect of heating treatment on the macro- and micro-structures

Fig. 3 shows the macrostructures of the ingot billets after heat-treated consisting of seven parts from the top to bottom of the ingot. No clear beta flecks or high density inclusions were found in different positions. There is no obvious change of grain shape zone with the ingot before heat-treating (Fig. 3(a)).

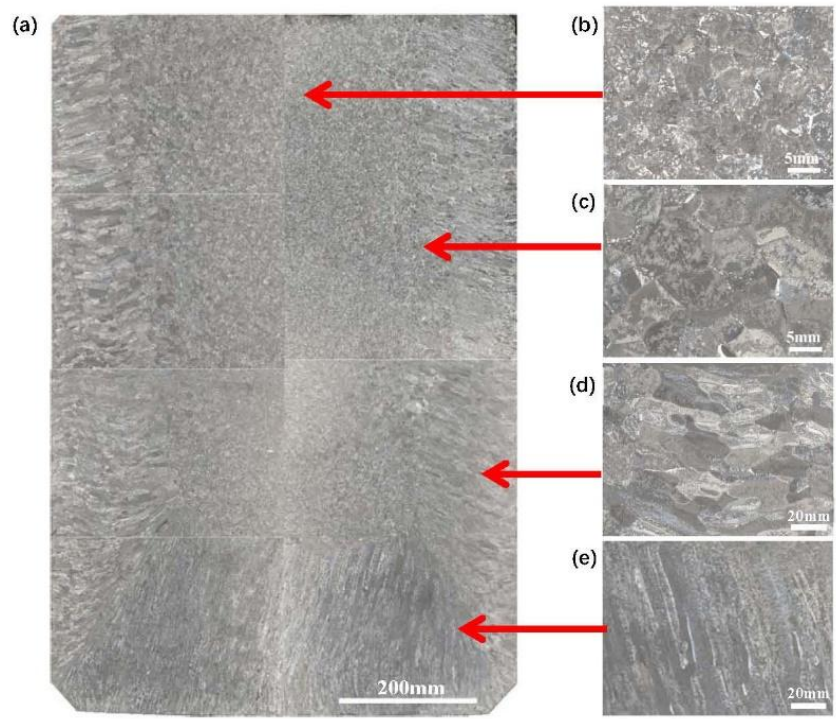

Fig. 3 macrostructures of the bilet along the longitudinal section after 940 ℃/1 h/WQ: (a) macrostructure; (b) equiaxed grain in the center; (c) equiaxed grain near the columnar grain zone; (d) columnar grain in the side; (e) columnar grain in the bottom.

Fig. 3(b) and 3(c) show the microstructures of the equiaxed grain in the center and near the columnar grain zone, respectively. The diameters of the equiaxed grain in the center were between about 3 mm and 5 mm，and the diameters of the equiaxed grain near the columnar grain zone were about 6–10 mm. Fig. 3(d) and 3(e) show the microstructures of the columnar grain in the side and the bottom, respectively. There are some differences between the columnar grain in the side and the bottom. The columnar grain in the side was in the shapes of "short stick", whose lengths were about 15–45 mm. The columnar grain in the bottom was in the shapes of "long stick", whose lengths were about 30–80 mm. The columnar grains are grown uniformly toward the hot top in the opposite direction of the temperature gradient.

Fig. 4 shows the microstructures of the equiaxed grain in the center, equiaxed grain near the columnar grain zone and the columnar grain in the side. The microstructures after heat-treated were the duplex microstructures with the β matrix and the α lamellar. The volume fraction of α phase is about 20%, which has small differences in different zones. The shapes of α phase were different with the length-width ratio, which increases from the center to the side of the ingot.

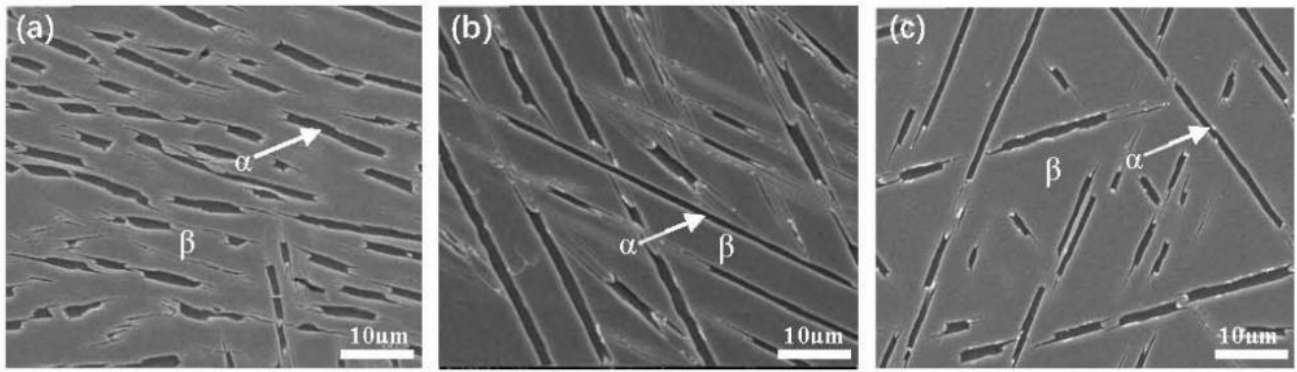

Fig. 4 Microstructures of grain in different sections after 940℃/1h/WQ: (a) equiaxed grain in the center; (b) equiaxed grain near the columnar grain zone; (c) columnar grain in the side.

Lots of micro-shrinkage cavities with the size of tens to hundreds of microns were found in the different positions of the ingot (Fig. 5). The shrinkage cavities are concave, which were formed in the solidification process because of the volume difference between the liquid and solidity. These zones can be recognized as the terminal of the solidification process, which also was the final solidification region of the interdendritic liquid phase [11,12]. Fig. 5 shows the microstructures near the shrinkage cavities. The microstructures and volume fractions of α phase were the same as the matrix structures. No clear beta flecks or high density enrichments were observed near the shrinkage cavities.

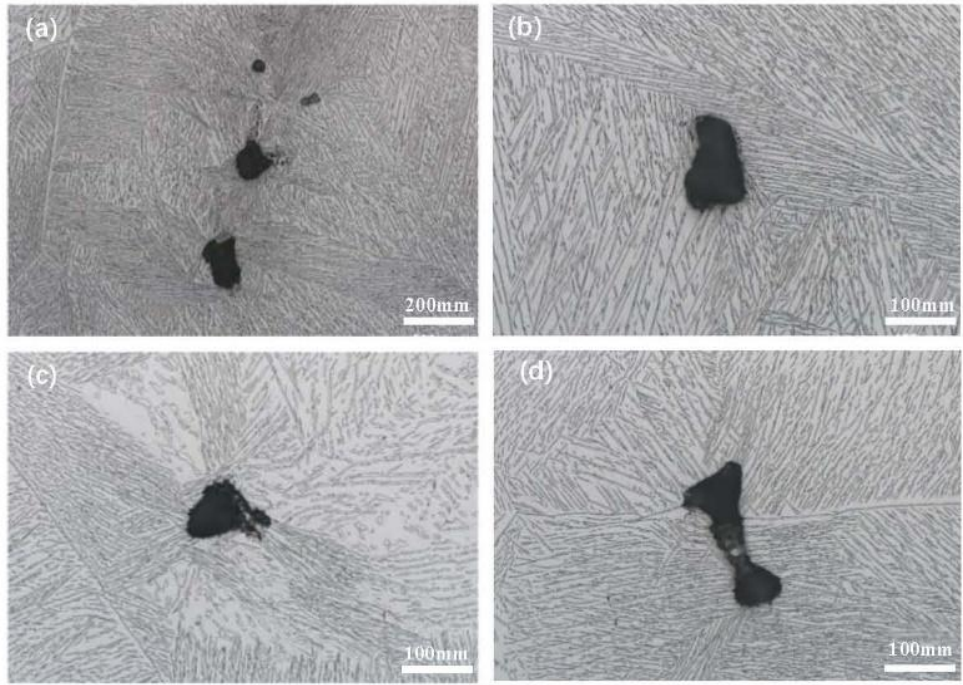

Fig. 5 microstructures near the micro-shrinkage cavities

3.3 Composition distribution in the industrial-scale ingot

Fig. 6 shows the typical curves of Al, Sn, Zr and Mo distribution in a cross-section from the top to bottom of the ingot. The curves show that Al and Sn compositions keep constant from the top to the bottom. The deviations of Al and Sn chemical compositions are very small. The minimum of Al is 5.94% and maximum is 6.12%, (the sample range is 0.18% only). The minimum of Sn is 2.0% and the maximum is 2.1%(the sample range is 0.1% only). The segregation degrees of Al and Sn are different from the results in Ref. [7]. Zr shows a normal composition in the round and bottom of the ingot but increases seriously to the maximum (4.48%) near the two shrinkage cavities in the hot top, where Mo decreases seriously to the minimum (5.51%). Zr decreases sharply to the minimum (3.55%) in the center of the ingot, where Mo increases to the almost maximum (5.95%) slightly. As can be seen, Mo exhibits an opposite trend with Zr. The more Zr, the less Mo.

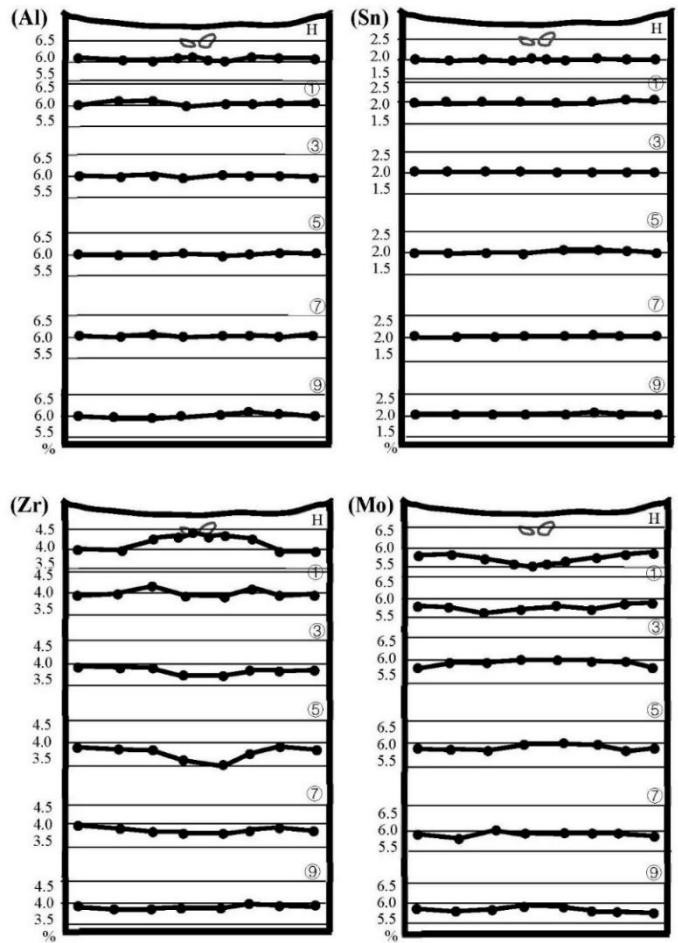

Fig. 6 Changes of Al, Sn Zr and Mo contents in cross-sections of the ingot

To illustrate the segregations area of elements more clearly, Fig. 7 shows the schematics of the Zr compositions in the ingot. The compositions on the surface usually are seen as the commercial chemical composition of the ingot. The average of Zr composition on the surface is about 3.88%, so 3.76%–4.0%(3.88% ±0.12%) can be seen in the regular range; 3.64%–3.75% and 4.01%–4.12% are acceptable as the low or high range separately. <3.64% and >4.12% are the serious low or high range unacceptably. It can be seen from Fig. 7 that Zr decreases sharply in the center of the ingot where the equiaxed grain zone is and increases seriously in the top of the ingot where the two shrinkage cavities are.

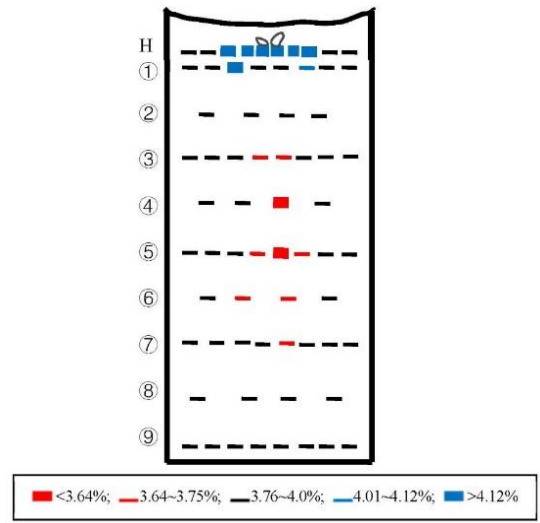

Fig. 7 Schematics of the Zr compositions in the longitudinal section of the ingot

## 4. Discussion

4.1 Solidified mechanism of marco-structures and micro-shrinkage cavities

The grain structure of VAR ingots may vary from almost completely columnar for small diameter ingots, to a mixture of columnar grains near the crucible walls and equiaxed grains in the central part of the ingot. The size of the central equiaxed region largely depends on the diameter of the ingot, the alloy properties, and the arc current, which is related to the intensity of the fluid flow in the sump [9]. As can be seen from Fig. 2, the columnar grain formed along the opposite direction of the temperature gradient in the industrial ingot. According to the VAR process, there is a bigger water flu in the bottom than in the round, the temperature gradient is also bigger so the columnar grain was in the shapes of "long stick"; the columnar grain was in the shapes of "short stick" because of the smaller temperature gradient in the round. There is an obvious phenomenon that almost all the columnar grain in the round grows toward two shrinkage cavities in the hot top. It means that temperature gradient is not to the center of the ingot horizontally, but the last solidified zone on the top of the ingot at an angle.

The micro-shrinkage cavities were found in the metal casting and melting process. They are usually considered to be caused by the density difference between the solid phase and the liquid phase formed during metal solidification. Some small volumes of liquid space are usually formed at the end of solidification, and the space is relatively closed. The fluidity of the liquid phase is poor, and density difference results in insufficient liquid feeding at the end of solidification. Finally, the solid volume shrinks to form pores[11–13]. No clear microstructure differences were found near the micro-shrinkage cavities, so it means that there are no obvious elements of micro-segregation such as high melting point Mo at the end of solidification.

4.2 Segregation mechanism of Zr and Mo

The macrosegregation in the columnar grain zone is slighter than that in the equiaxed grain zone and hot top zone because of the higher temperature gradient. The electromagnetic stirring accelerated the melt flow in the molten pool, the strong melt flow broke the correspondence between the grain structures and promoted the transition from the the columnar grain to equiaxed grain during the VAR process. It means that the electromagnetic stirring decreased the temperature gradient in the center. The center segregation in the equiaxed grain zone (Fig. 7) had not been improved by electromagnetic stirring. The result is different from Ref. [14] which shows that adding magnetic stirring and adoptting small melting current decreases the macrosegregation rate of Fe element in Ti-10V-2Fe-3Al. The macrosegregation in the center of the Ti-1023 alloy ingot was more serious when the melting current increases [9,14], so small melting current must be the main factor on the decrease of macrosegregation rate of Fe element in the Ti-1023 titanium alloy but the electromagnetic stirring. Furthermore, the results of Ref. [15] also show that the center segregation was effected by electromagnetic stirring parameters such as current intensity and stirring pool width. Large ingot size will increase the degree of macrosegregation [16], electromagnetic stirring can effect the Lorentz force distribution inside the metal on the liquid pool in the VAR process and deformation of the free surface of the bath [17]. It could be concluded that the temperature gradient and solidification rate in the industrial-scale ingot dominated the macrosegregation formations during the solidification process.

In consideration of the interaction between elements, the distribution of solute elements in the solid-liquid phase was calculated by Jmatpro software (Fig. 8). The contents of Zr and Sn in the liquid increase at the solid-liquid transformation but Al and Mo exhibit an opposite trend. It means that Zr and Sn enriched in liquid, resulting in high contents at the end of solidification. Al and Mo poored in liquid, resulting in low contents at the end of solidification. It is not much consistent with the analysis results from Fig. 6, in which Al and Sn keep constant always but Zr and Mo show the macrosegregation in the ingot. It means that the results calculated by software do not represent the actual results.

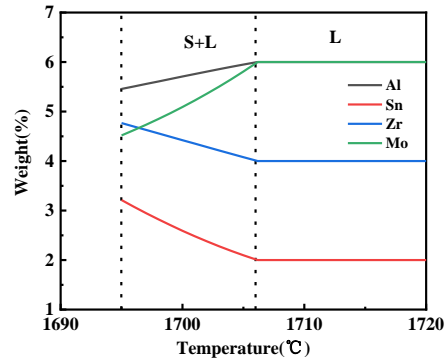

Fig.8 Distribution of elements in liquid at solid-liquid transformation of Ti6246 alloy

The partition coefficient ($k$) of the alloy elements represents the segregation degree of each element in the solidification of alloys [18]. When $k=1$, there is no segregation of solute atoms at the solid-liquid interface; when $k>1$; solute atoms are preferentially enriched in the solid phase at the solid-liquid interface and form negative segregation; when $k<1$, solute atoms are preferentially enriched in the liquid phase at the solid-liquid interface and form positive segregation.

Table 2 Partition coefficients (k) of alloying elements in titanium alloys[18]

| Al | Sn | Zr | Mo | V | Cr | Fe | Alloys |
|---|---|---|---|---|---|---|---|
| 1.05 | 0.92 | 0.90 | 1.50 | 0.90 | 0.77 | 0.60 | Binary phase diagram |
| 1.13 | - | - | - | 0.95 | - | 0.38 | Ti-1023 |
| 1.06 | 1.15 | 0.77 | 1.15 | - | 0.74 | - | Ti-17 |
| 1.02 | 1.08 | 0.72 | 1.09 | - | - | - | Ti-6242 |

The main β-phase stabilizing elements are Mo, Cr, V or Fe in the β-Ti alloys such as Ti-17 and Ti-1023 alloys. Elemental segregation is visualized by areas with dark and light contrasts in the macrostructure of Ti-17alloy. Dark contrast areas are rich of Cr ($k<1$), Zr ($k<1$) but lean of Mo ($k>1$)

[10]. According to Table 2, Cr ($k$=0.74–0.77) and Fe ($k$=0.38–0.60) are the serious positive segregation in the later solidification. Al ($k$=1.02–1.13) and Sn ($k$=0.92–1.15) are not the segregation in the solidification because $k$ are nearly close to 1. Zr ($k$=0.72–0.90) is the negative segregation and Mo ($k$=1.09–1.50) is positive segregation in the solidification. It shows the same situation in this paper. So Al and Sn are not the segregation in the solidification in this work because $k$ is nearly close to 1. Zr is the negative segregation and Mo is the positive segregation in the solidification because $k$ is far away from 1. As the positive segregation, Mo is enriched in the solid phase at the solid-liquid interface. The volume fraction of the solid phase is higher than that of the liquid phase at the end of the solidification so that Mo is fully diluted in the liquid. Furthermore, the diffusion and fluidity of the solid phase are much lower than that of the liquid phase so that it is hard to form high concentration enrichment of Mo in the micro-area. This is the reason why no clear beta flecks or high density enrichments were found near the shrinkage cavities (Fig. 5). Zr is the negative segregation in the solidification and enriched in the last solidification zones. However, the content of Zr in the round of ingot is higher than that in the central equiaxed region of the ingot. It means that the central equiaxed region is not the last solidified zone. The contents of Zr near the two shrinkage cavities in the hot top are the most because the last solidified zone is on the top of the ingot which is in accord with the macrostructure of the ingot (Fig. 2).

The β-flacks formed in the center of the ingot are due to the falling of equiaxed grains, which is called "crystal rain" [19]. The volume of the liquid metal poll is large when the melting rate is higher of VAR, which is up to 60%–70% of the total ingot in some titanium alloy ingots [8]. At this time, the solidification process is closer to the common large volume casting process. During solidification, the accumulated crystals capture the segregated liquid phase in the molten pool and fix the liquid phase in the crystal space, resulting in the formation of β-flacks in this region. The equiaxed crystal is nucleated in the molten pool when the molten pool is characterized by low circulating flow and low temperature gradient [20]. In Ti alloys, the solidified crystal sinks to the bottom of the molten pool in the form of a raindrop because the density is higher than that of liquid. This kind of equiaxed solidification segregation often occurs in alloy solidification with a large melting pool, slow cooling rate and low temperature gradient [21,22]. In this work, the Ti-6246 alloy industrial-scale ingot has the same big volume of the liquid metal poll, low circulating flow and low temperature gradient. It means that the solidified equiaxed crystal sinks to the bottom of the molten pool in the form of a raindrop. However, the accumulated crystals did not capture the segregated liquid phase depleting Zr in the molten pool and did not fix the liquid phase in the crystal space during solidification.To verify the above viewpoint, the density change of Ti6246 alloy during solid-liquid phase transformation was calculated by Jmatpro software (Fig. 9). With the solidification process, the liquid density of Ti6246 alloy increases gradually. The density of the liquid phase is about 0.213 g/cm$^3$ lower than that of the solid phase. This density difference is enough to drive the solid phase to sink in the equiaxed region of the ingot center.

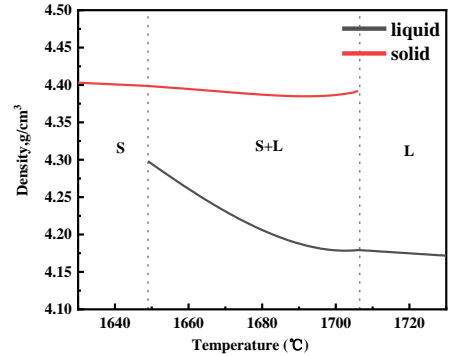

Fig. 9 Density of Ti6246 alloy calculated by Jmatpro software at 1630–1730 ℃

The natural buoyancy of the molten pool can promote the macrosegregation in VAR progress, and the segregation solute is driven to the center of the molten by the natural buoyancy in the form of heat flow. The liquid flow formed a stable density flow. The thermal buoyancy makes the solute element negative segregation Zr is continuously discharged to the front of the solid-liquid interface. Because positive segregation Mo is enriched in the solid, the density of the liquid in the front of the solid-liquid interface is smaller than that of the solid. Thus, the upward solute buoyancy is formed. Under the combined action of thermal buoyancy and solute buoyancy, the melt convection is formed in the molten pool with the solid phase downward and the liquid phase upward. This convection brings the Zr from the solid-liquid interface to the upper part of the ingot. The results show that the

Zr in the equiaxed crystal zone in the center of the ingot is reduced, and the Zr is enriched in the upper riser as the end of solidification.

**5. Conclusion**

1. Ti6246 titanium alloy industrial-scale ingot by VAR is characterized by columnar grain zone, equiaxed grain zone and hot top zone. Almost all the columnar grain in the round grows toward two shrinkage cavities in the hot top. No clear beta flecks or high density enrichments were observed near the shrinkage cavities.

2. The macrosegregation in the columnar grain zone is slighter than that in the equiaxed grain zone because of the higher cooling rate, but the center segregation could not be improved more obviously by electromagnetic stirring because of center liquid solute enrichment and liquid phase accumulation in the stirring zone.

3. As the positive segregation, Mo is enriched in the solid phase at the solid-liquid interface. The volume fraction of the solid phase is higher than that of the liquid phase at the end of the solidification so that Mo is fully diluted in the liquid. Furthermore, the diffusion and fluidity of the solid phase are much lower than that of the liquid phase so that it is hard to form high concentration enrichment of Mo in the micro-area. Zr is the negative segregation in the solidification and enriched in the last solidification zones. The thermal buoyancy made the negative segregation Zr be continuously discharged to the front of the solid-liquid interface.

References

[1] R. R. Boyer, An overview on the use of titanium in the aerospace industry, Mater. Sci. Eng. A 213 (1996) 103-114.

[2] P. Wanjara, M. Jahazi, H. Monajati, S. Yue, Influence of thermomechanical processing on microstructural evolution in near-α alloy IMI834, Mater. Sci. Eng. A 416 (2006) 300–311.

[3] Paul J. BANIA, Next Generation Titanium Alloys for Elevated Temperature Service, ISIJ International 31(8)

(1991) 840-847.

[4] Weiju Jia, Weidong Zeng, Yigang Zhou, Jianrong Liu, Qingjiang Wang, High-temperature deformation behavior of Ti60 titanium alloy. Mater. Sci. Eng. A 528 (2011) 4068-4074.

[5] A. Mitchell, Melting, casting and forging problems in titanium alloys, Mater. Sci. Eng. A 243 (1998) 257-262.

[6] V.I. Dobatkin, N.F. Anoshkin, Comparison of macrosegregation in titanium and aluminium alloy ingots, Mater. Sci. Eng. A 263 (1999) 224-229.

[7] Xin Liu, Guang Feng, Yu Zhou, Qunbo Fan, Macrosegregation and the underlying mechanism in Ti-6.5Al-1.0Cr-0.5Fe-6.0Mo-3.0Sn-4.0Zr alloy, Progress in Natural Science: Materials International 29 (2019) 224-230.

[8] A. Mitchell, Solidification in remelting processes, Mater. Sci. Eng. A 413-414 (2005) 10-18.

[9] Dmytro V. Zagrebelnyy, Modeling macrosegregation during the vacuum arc remelting of Ti-10V-2Fe-3Al Alloy, Ph.D. Thesis PURDUE University (2007).

[10] Xuchen Yin, Jianrong Liu, Qingjiang Wang, Lei Wang, Investigation of beta fleck formation in Ti-17 alloy by directional solidification method, J. Mater. Sci. Technol. 48 (2020) 36-43.

[11] P.D. Lee, R.C. Atwood, R.J. Dashwood, H. Nagaumi, Modeling of porosity formation in direct chill cast aluminum-magnesium alloys, Mater. Sci. Eng. A 328 (2002) 213-222.

[12] Giulio Timelli, Daniele Caliari, Jovid Rakhmonov, Influence of Process Parameters and Sr Addition on the Microstructure and Casting Defects of LPDC A356 Alloy for Engine Blocks, J. Mater. Sci. Technol. 32 (2016) 515-523.

[13] Zhiming Gao, Wanqi Jie, Yongqin Liu, Yongjian Zheng, Haijun Luo, A model for coupling prediction of inverse segregation and porosity for up-vertical unidirectional solidification of Al-Cu alloys, J. Alloys. Compou. 797 (2019) 514-522.

[14] Zhijun Yang, Hongchao Kou, Jinshan Li, Rui Hu, Hui Chang, Lian Zhou, Macrosegregation Behavior of Ti-10V-2Fe-3Al Alloy During Vacuum Consumable Arc Remelting Process, J. Mater. Eng. Perform. 20(1) (2011) 65-70.

[15] Dongbin JIANG and Miaoyong ZHU, Center Segregation with Final Electromagnetic Stirring in Billet

Continuous Casting Process, Metal. Mater. Trans. B 48B (2017) 444-455.

[16] Dmytro Zagrebelnyy, Matthew John M. Krane, Segregation Development in Multiple Melt Vacuum Arc Remelting. Metal. Mater. Trans. B 40B (2009) 281-288.

[17] P Chapelle, A Jardy, J.P. Bellot, Effect of electromagnetic stirring on melt pool free surface dynamics during vacuum arc remelting[J]. J. Mater. Sci. 43(17) ( 2008) 5734-5746.

[18] A. Mitchell, A. Kawakami, S.L. Cockcroft, Segregation in titanium alloy ingots, High Temper. Mater. Process. 26(1) (2007) 59-77.

[19] A. Mitchell, A. Kawakami, S.L. Cockcroft, Beta fleck and segregation in titanium alloy ingots, High Temper. Mater. Process. 25(5-6) (2006) 337-349.

[20] Kurz W, Fisher D J. Fundamentals of solidification[M]. Zurich: Trans Tech Publications, 1998.

[21] Jun Li, Menghuai Wu, Andreas Ludwig, Abdellah Kharicha, Simulation of macrosegregation in a 2.45-ton steel ingot using a three-phase mixed columnar-equiaxed model, Inter. J. Heat. Mass. Trans. 72 (2014) 668-679.

[22] J. A. Spittle, Columnar to equiaxed grain transition in as solidified alloys. Inter. Mater. Review. 51(4) (2006) 247-269.